\documentclass[10pt,conference]{IEEEtran}
\IEEEoverridecommandlockouts

\usepackage{cite}
\usepackage{amsmath,amssymb,amsfonts}
\usepackage{algorithmic}
\usepackage{graphicx}
\usepackage{textcomp}
\usepackage{xcolor}
\usepackage{graphicx} 
\usepackage{tikz}
\usepackage{caption}
\usepackage{subcaption}
\usepackage{siunitx}
\usepackage{textcomp}

\usepackage{flushend}
\usetikzlibrary{shapes.geometric, arrows, positioning}

\begin{document}

\title{Handling Large-Scale Network Flow Records:\\A Comparative Study on Lossy Compression}

\author{
    \IEEEauthorblockN{
    Gabriele Merlach\IEEEauthorrefmark{1},
    Martino Trevisan\IEEEauthorrefmark{1},
    Damiano Ravalico\IEEEauthorrefmark{1}, \\
    Fabio Palmese\IEEEauthorrefmark{2},
    Giovanni Baccichet\IEEEauthorrefmark{2},
    Alessandro Redondi\IEEEauthorrefmark{2}}
    
\IEEEauthorblockA{\IEEEauthorrefmark{1}University of Trieste, \IEEEauthorrefmark{2}Politecnico di Milano} }

\maketitle

\begin{abstract}
Flow records, that summarize the characteristics of traffic flows, represent a practical and powerful way to monitor a network. While they already offer significant compression compared to full packet captures, their sheer volume remains daunting, especially for large Internet Service Providers (ISPs). In this paper, we investigate several lossy compression techniques to further reduce storage requirements while preserving the utility of flow records for key tasks, such as predicting the domain name of contacted servers. Our study evaluates scalar quantization, Principal Component Analysis (PCA), and vector quantization, applied to a real-world dataset from an operational campus network. Results reveal that scalar quantization provides the best tradeoff between compression and accuracy. PCA can preserve predictive accuracy but hampers subsequent entropic compression, and while vector quantization shows promise, it struggles with scalability due to the high-dimensional nature of the data. These findings result in practical strategies for optimizing flow record storage in large-scale monitoring scenarios.
\end{abstract}


\section{Introduction}
\label{sec:intro}

Network monitoring has become increasingly essential as the complexity and scale of modern networks continue to grow. One crucial technique is \emph{passive monitoring}, wherein a passive meter continuously observes and records the traffic generated by a population of users or servers. Rather than capturing the entire packet stream, passive meters summarize traffic in the form of flow records. These flow records, typically generated for every TCP or UDP connection, include features such as packet size, timing information, or metadata derived from Deep Packet Inspection, such as domain names. Passive monitoring serves a wide range of applications and is extensively used by Internet Service Providers (ISPs) and network administrators for traffic accounting, quality of service (QoS) monitoring, and the detection or investigation of cybersecurity-related events.

One of the primary benefits of flow records is their significant data reduction compared to full packet capture. Specifically, flow records provide over $1000$ times compression relative to the size of a complete packet trace, as highlighted by Hofstede et al.\cite{hofstede2014flow}. Nonetheless, full packet traces are still required for certain applications, such as forensic investigations, and several effective network trace compression strategies have been proposed in the literature~\cite{holanda2005performance,liu2005information,chen2008ipzip}.

Despite the compression benefits provided by flow records, their storage can become a considerable challenge in large-scale monitoring environments. For instance, monitoring the traffic generated by \num{10000} households can result in approximately \SI{500}{\giga\byte} of flow records per month~\cite{trevisan2018five}. In larger settings, such as an Internet Service Provider or a Satellite Communication (SatCom) operator, this figure can grow exponentially. For example, flow records generated by a SatCom operator may amount to around $10$ billion records per month, requiring up to \SI{4}{\tera\byte} of storage for compressed logs~\cite{perdices2022satellite}. Flow records are typically stored using well-known standards like CSV files and Parquet tables, or injected into databases designed for large-scale data processing, such as HBase or Hive from the Hadoop ecosystem. While these systems support traditional entropic compression algorithms, they generally lack specialized compression schemes tailored for flow records. To address this, prior research has explored \emph{lossless} compression strategies customized for flow data. These strategies include bitmap schemes~\cite{fusco2010net}, encoding techniques specifically designed for network traffic data~\cite{fusco2012rasterzip}, and pattern classification methods~\cite{cornelisse2016compressing}.

In this work, we evaluate various lossy compression schemes for flow records, considering the scenario where a network administrator or ISP needs to store many of them for a specific need over extended periods while minimizing the required storage. Going beyond traditional entropic compression, we assess how different straightforward lossy compression algorithms can reduce data size while maintaining acceptable accuracy of flow record analysis. 
Specifically, we focus on a classification task where flow records are used to predict the contacted server's domain name (i.e., the website). We study the trade-off between compression ratio and classification accuracy. Although our methodology considers this specific use case, it can easily be extended to other scenarios.

We explore several lossy compression techniques using a large-scale dataset collected from an operational campus network. Our results can be summarized as follows:
\begin{itemize}
    \item Scalar quantization, despite its simplicity, offers the best trade-offs between compression ratio and data accuracy.
    \item Principal Component Analysis maintains good accuracy if most of the original variance is preserved. However, due to the inherent sparseness of flow records, an increase in signal entropy in the PCA domain is observed, which jeopardizes compression efficiency.

    \item Vector quantization, which we implement using K-Means, shows promising performance but struggles to scale with high-dimensional large datasets such as the one used in this work, limiting practical application.
\end{itemize}

\section{Methodology}
\label{sec:metho}

In our experiments, we evaluate lossy compression methods aimed at reducing the size of flow records. Each record represents a network flow (a TCP or UDP connection), identified by its 5-tuple, i.e., source and destination IP address and port, and L4 protocol employed (TCP or UDP). A record, besides the 5-tuple, carries various metrics describing the flow. The set of metrics heavily depends on the employed flow exporter and may include: the number of exchanged bytes and packets in both directions, performance metrics (such as Round Trip Time, 
or bitrate), the size and timings of the first few packets and information extracted through Deep Packet Inspection (DPI), such as domain names, TLS certificates or HTTP headers. The following lossy compression schemes are evaluated:

\subsection{Scalar Quantization}

Scalar quantization (SQ) is a fundamental technique in signal processing and data compression, wherein continuous data samples are normalized and subsequently quantized within a specified range using $B$ bits per sample. Here, each flow record metric is quantized independently with $B$ denoting the number of bits allocated for the representation of each quantized value of the columns. The quantized value  $v_i^\prime$ is computed using the following formula:

\begin{equation*}
    v_i\prime = \left\lfloor \frac{\min(\max(v_i, P_{1,i}), P_{99,i}) - P_{1,i}}{P_{99,i} - P_{1,i}} \cdot (2^B - 1) \right\rceil 
\end{equation*}
    
Employing percentiles in the quantization process allows to mitigate the impact of outliers on the quantizer design, leading to more robust results compared to using minimum and maximum values. In our work, we always use the $1$st and the $99$th percentiles instead of minimum and maximum and evaluate the impact of the value of the bit depth $B$, where, intuitively, a larger $B$ maximize accuracy, while a smaller $B$ maximizes compression.

\subsection{Principal Component Analysis}

Principal Component Analysis (PCA) is a statistical technique used for dimensionality reduction. It operates by decreasing the number of features in the data frame while preserving as much of the original variance as possible. It is based on the decomposition of the covariance matrix of the data, leading to the identification of eigenvalues and eigenvectors.
The PCA process results in the storage of key components, including the eigenvalues, the eigenvectors (which represent the principal components), and the transformed data coordinates. This information enables the approximate reconstruction of the original data from the principal components. However, it is important to note that the quality of the reconstruction is contingent upon the amount of variance captured by the selected components.

In this work, we use PCA as a compression technique. We train a PCA model on the data and store (i) the key components and (ii) the transformed data, so that it is possible to map back data on the original space -- effectively operating a decompression. We evaluate PCA  for different levels of variance retained by key components. As the retained variance decreases, the number of principal components is reduced, which leads to smaller expected file sizes.

\subsection{Vector Quantization}

Vector Quantization is a technique that maps continuous vectors to a discrete space by clustering similar data points. Each data point is quantized to the nearest cluster centroid, effectively reducing the dataset's complexity while preserving essential information. This process aids in data compression by representing similar values with a single representative vector per cluster. To implement this, we use K-Means clustering, relying on its improvement K-Means++ for centroid initialization. This enhances computational efficiency with large-scale datasets compared to random initialization~\cite{arthur2006k}. Notice that the computational complexity of K-Means is:
\begin{equation*}
    O(n \cdot K \cdot t \cdot d)
    \label{eq:kmeans_complexity}
\end{equation*}
where $K$ is the number of clusters, $n$ is the number of points in the dataset, $t$ number of iterations before convergence and $d$ is the dimensionality of the dataset. In our problem, most of those parameters are large, as we aim at compressing millions of flow records ($n$), each represented by more than one hundred features ($t$ -- see Section~\ref{sec:dataset}) and the large diversity of the data, i.e., flow records from thousands of devices to different servers, requires a high number of clusters ($K$).

As we are coping with large datasets and high-dimensional spaces, we evaluate the performance for different values of $K$. To further enhance scalability, we experiment with centroid computation on just a subset of the $n$ original points.
In our preliminary experiments, we also evaluated density-based clustering algorithms, i.e., DBSCAN and its hierarchical version HDBSCAN. However, even when using indexed data structure for fast neighbour lookup, their computational complexity remains $O(n \cdot log(n))$, which made the computational time too large for a practical application, being $n \sim 10^6$.

\subsection{Evaluation Methodology}

\begin{figure}[!t]
    \centering
    \includegraphics[width=1\columnwidth]{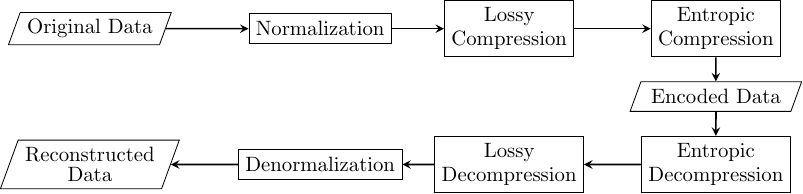} 
    \caption{Diagram of encoding-decoding process.}
    \label{fig:encoding_decoding}
\end{figure}

We evaluate the algorithms above to compress flow records and build compression-decompression pipelines. We test the three compression approaches alone and combine them to study potential mutual benefits. For each algorithm, we vary its main parameters to evaluate its efficacy and measure the tradeoffs between compression ratio and data utility. We sketch the building blocks of our compression-decompression pipelines in Figure~\ref{fig:encoding_decoding} and describe them below.

\noindent
\textbf{Data Normalization}.
The data is always normalized to achieve a zero median and unit variance, ensuring consistency in the analysis. After decompression, the encoded data is de-normalized as the final step in the reconstruction process, as shown in Figure~\ref{fig:encoding_decoding}, while preserving information about the normalization and quantization parameters in the resulting files. The dimensions of these de-normalization files are not considered relevant for evaluating the quantization process and are therefore omitted from further analysis.

\noindent
\textbf{Lossy Compression}.
At this step, data is compressed using one of the algorithms above or more than one, in cascade. Specifically, we evaluate:
\begin{itemize}
    \item \textbf{Scalar Quantization} (i) alone
    \item \textbf{PCA}: (i) alone and (ii) followed by Scalar Quantization to evaluate whether the transformed data can be further compressed
    \item \textbf{Vectorial Quantization} with (i) K-Means on the full dataset and (ii) using a sub-sample of the data to increase scalability 
\end{itemize}

\noindent
\textbf{Entropic Compression}.
Once the data has been transformed with one of the above techniques, we further compress them with Gzip to remove any remaining redundancy. Gzip was ultimately selected as the primary method due to its efficiency, and tests with other algorithms (Zip, RAR and 7z) do not lead to significantly different results. We underline that this entropic compression step is essential, as we first run lossy compression techniques that operate on a different basis.

\noindent
\textbf{Compression Ratio.} We aim to determine whether applying lossy quantization can effectively reduce the size of the files in our data lake. We compare the quantized compressed binary files with the original compressed binary files stored in the data lake. The metric used for this comparison is the Compression Ratio, defined as:
\begin{equation*}
\footnotesize
\text{Compression Ratio} = \frac{\text{Size with Lossy Compression \emph{and} Gzip}}{\text{Size with Gzip}} \times 100
\end{equation*}
A compression ratio of less than $100$ indicates a successful reduction in the size of the data stored in the datalake, which is the desired outcome. 

\noindent
\textbf{Data Utility.} In addition to reducing file sizes, it is crucial that the quantized data maintain performance levels comparable to the original. We evaluate data utility in a classification problem that we detail in Section~\ref{sec:dataset}.
\section{Use Case and Dataset}
\label{sec:dataset}

\subsection{Dataset}
\label{sec:data}

In our experiments, we use a flow record dataset from a University Campus during a full day. The data were captured using a passive probe located at the edge router of the campus, aggregating individual packets into flows through the network traffic analyzer Tstat~\cite{trevisan2017traffic}. Tstat is a passive traffic monitoring tool that generates a record for each observed TCP/UDP stream. Each record includes more than a hundred basic and advanced flow features such as IP addresses, port numbers, domain names extracted from TLS Server Name Indications (SNIs), and various statistics, including packet/byte counters, retransmission counts, and more. In particular, there exist $175$ numerical metrics that we aim at compressing and are used as features for the classification problem we present next. These include metrics like the size of the first $10$ packets, the number of retransmitted packets, and flow duration. All the non-numerical metrics, such as IP addresses or domain names, are not considered in the lossy compression process. To ensure user privacy, all IP addresses were anonymized immediately, preventing the identification of individual users. Tstat discarded any potential privacy-sensitive information, retaining only essential statistics and domain names. 

Over one day of traffic observation, more than $14.3$ million flows were recorded, involving over \num{3800} distinct client IP addresses communicating with approximately \num{380000} unique server IPs. The network logs are compressed using the Gzip entropy-based compressor and then stored in a data lake. The size of the Gzip-compressed logs for a single day of traffic is \SI{14.9}{\giga\byte}.\footnote{The entropy compressor already reduced the size by $84$\%.}

\subsection{Use Case: Domain Classification}
\label{sec:classification}

We quantify the trade-offs between lossy compression and data utility with a simple classification task on flow records. We take this case study as classification makes use of all numerical features of flow records, being thus an effective way to quantify to what extent compressed data can serve the same way as original data.

The classification problem we target is the same as in our previous works~\cite{trevisan2020does,trevisan2023attacking}. Indeed, one of the purposes of network data collection is the investigation of the efficacy of encryption methods. Specifically, we use the data to train a machine learning model designed to classify the domains contacted by clients when such information is obscured. Random Forest has been previously demonstrated to be the most effective method for this classification problem, owing to its simplicity and efficacy.

We focus exclusively on domains with a minimum of \num{1000} traffic flows, hosted by widely recognized Autonomous Systems (AS). This selection results in 35 ASes, which include top content providers, Content Delivery Networks (CDN), and cloud service providers, collectively accounting for $90$\% of the total observed traffic flows.
We employ the SNI of HTTPS flows to label the domains contacted by the clients. Non-popular domains are categorized under the label ``Other'', representing an imbalanced class containing flows with diverse feature characteristics.

Given the substantial number of total classes -- $708$ distinct classes -- we train a separate Random Forest model for each AS, utilizing only the traffic flows associated with that specific AS. This choice fits our classification problem, as the server IP address can be easily mapped to the AS, thus restricting the set of possible domains to those hosted in the AS. The dataset is partitioned into training and testing sets, randomly selecting $50$\% of the client IP addresses for training the Random Forest model, with the remaining $50$\% reserved for testing. The performance of the Random Forest model is evaluated by calculating the F1-Score for each class within every AS, where an F1-Score of $1$ indicates perfect classification on the test set.

\section{Results}
\label{sec:results}

In this section, we report the results of our experiments, evaluating the different compression strategies. For each method, we vary the most important parameters and quantify results in terms of compression ratio and F1-Score.  We run experiments on a server equipped with \SI{192}{\giga\byte} RAM and two Intel\textsuperscript{\textregistered} Xeon\textsuperscript{\textregistered} Gold 6140 CPUs @\SI{2.30}{\giga\hertz} with $18$ cores and $36$ threads. Data are stored on a distributed file system based on Ceph.

\subsection{Scalar Quantization}

\begin{figure}
    \centering
    \includegraphics[width=\linewidth]{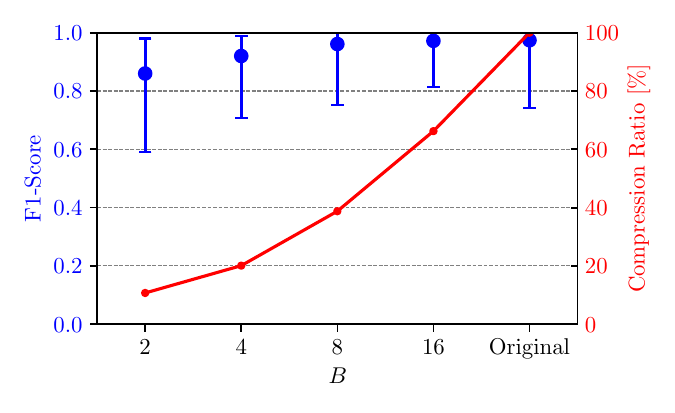}
    \vspace{-5mm}
    \caption{Compression ratio and F1-Score with Scalar Quantization with different compression depth $B$.}
    \label{fig:scalar}
    \vspace{-5mm}
\end{figure}

We first focus on scalar quantization, which, as anticipated offers good results both in terms of compression ratio and F1-Score. Figure \ref{fig:scalar} illustrates the trade-offs between compression ratio and data utility at varying bit depths, denoted by $B$. The red curve (right $y$-axis) shows the compression ratio, which increases as the bit depth decreases, reaching $11$\% when using as low as $2$ bits per record field.\footnote{We remark that compression ratio compares the size of (i) data compressed using lossy compression and entropic compression with (ii) data compressed uniquely with entropic compression.} Notice that original data uses $32$ bit integer and floating point fields, making, thus, compression ratio substantial when reducing the field granularity. Observe, however, that required storage does not scale linearly with $B$, as we use a final entropic compression stage, which reduces redundancy in the data regardless of the number of bits used to represent quantities.

The blue points (left $y$-axis) represent the F1-Score obtained on the domain classification task, indicating the performance of classification across different compression levels. Each blue point represents the median F1-Score across multiple domains ($708$ in our case), with error bars extending from the $5$th to the $95$th percentiles, quantifying the variability in classification performance across these domains. The plot demonstrates a clear trade-off: as the compression ratio improves with reduced bit depth, there is a corresponding decrease in the F1-Score, quantifying how higher compression ratios compromise the utility of the data for classification tasks. However, we observe a remarkable drop in performance only when $B=2$, leading to a median F1-Score of $0.86$. Already with $B=4$, the median F1-Score is $0.92$, and we achieve a compression ratio of $21$\%, meaning that storage size is reduced by $5$ times. Notice that in this case F1-Score is reduced only by $0.07$ with respect to classification on the original data. The time requested to compress the entire dataset with SQ is 12 minutes, regardless of the bit depth.

\subsection{Principal Component Analysis}

\begin{figure}
    \centering
    \begin{subfigure}[b]{\columnwidth}
        \centering
        \includegraphics[width=0.8\linewidth]{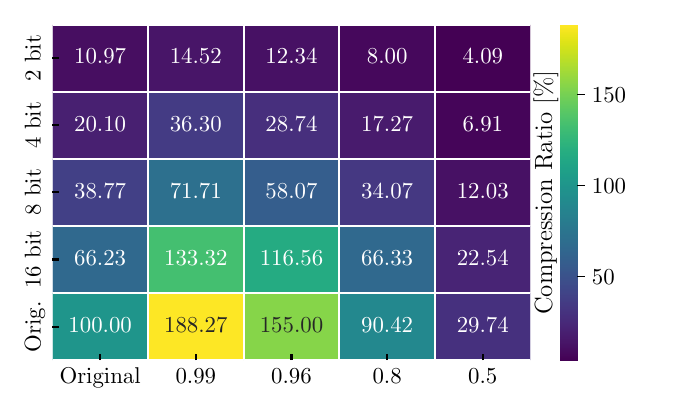}
        \caption{Compression Ratio}
        \label{fig:pca-compression}
    \end{subfigure}
    \begin{subfigure}[b]{\columnwidth}
        \centering
        \includegraphics[width=0.8\linewidth]{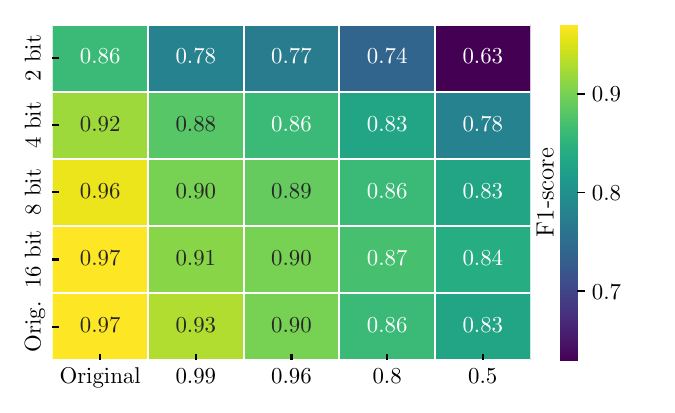}
        \caption{Classification Performance}
        \label{fig:pca-classification}       
    \end{subfigure}

    \caption{Compression ratio and Classification Performance with PCA and Scalar Quantization.}
    \label{fig:pca}
\end{figure}

We now focus on PCA, using it in isolation and in combination with scalar quantization. In the latter case, we first run PCA and then apply scalar quantization to the data represented with the identified components. We report results in Figure~\ref{fig:pca} in the form of a heatmap, where each cell represents the value obtained using PCA when preserving a certain amount of the original variance (different columns) and when applying a certain amount of scalar quantization (different rows). With PCA, we do not fix a number of principal components but set their number to reach the desired level of original variance, allowing fair comparison between different ASes. Indeed, the number of resulting components differs for each AS, as we compress data independently per AS. Overall, when preserving $0.99$, $0.96$, $0.8$, $0.5$ of the original variance, on average, we obtain $115$, $100$, $80$, and $15$ components, respectively. 

Starting from the compression ratio (Figure~\ref{fig:pca-compression}), we first remark that, as we decrease the bit depth of scalar quantization from the original $32$-bit to $2$-bit, the compression ratio consistently improves (lower values are better), as already shown by Figure~\ref{fig:scalar}. This trend holds when applying PCA together with scalar quantization (all columns except left-most), but PCA drastically impacts compression performance, particularly when preserving most variance. Indeed, at $0.99$ variance retention with no quantization ($32$-bit), the compression ratio jumps to $188.27$, suggesting that PCA injects substantial entropy into the data, making it harder to compress effectively. 
This occurs because PCA redistributes the variance across the principal component, thereby reducing the correlation and redundancy in the data. As a result, the Gzip compression algorithm, which is optimized for exploring data redundancy and patterns of repetitions, became less effective due to the increased entropy and reduced predictability of other transformed data. Even at $16$-bit, the $0.99$ PCA retention level yields a high compression ratio of $133.32$, while dropping to $2$-bit with $0.5$ retention reduces it to $4.09$. This is particularly interesting if compared to the $10.97$ compression ratio with $2$-bit and no PCA, which operates on the original $175$ dimensions, while only $15$ are left with $0.5$ PCA retention rate.

We investigate this phenomenon with Table~\ref{tab:entropy_tab}: The original dataset presents a median entropy of $8.37$ and a decompressed size of $611.40$\% with respect to the Gzip compressed version ($100$\%, our benchmark), reflecting its structural redundancy. In contrast, when using PCA at $0.99$ variance, the median entropy increases to $22.56$ and non-Gzip compressed size decreases to $219.28$\%. However, Gzip efficacy is minimal, reducing data size only to $188.27$\%, highlighting how the increased entropy leads to diminished compression effectiveness. Interestingly, with $0.5$ variance retention, despite a significant feature reduction to $15$, the median entropy remains relatively high at $19.56$ and the effective compression ratio is just $29.74$\% suggesting that while PCA has reduced the overall data volume, the high entropy still presents a challenge for compression.

Focusing now on classification performance, Figure~\ref{fig:pca-classification} shows the median F1-Score across the $708$ domains, indicating how well the classification task can be performed after compression. As shown in Figure~\ref{fig:scalar}, applying scalar quantization deteriorates classification performance as bit depth decreases. Similar considerations hold with PCA: Decreasing variance retention levels leads to a drop in F1-Score. For instance, at $16$-bit quantization, the F1-Score decreases from $0.97$ to $0.91$ for $0.99$ retention and drops to $0.84$ at $0.5$ retention. We can make a clean comparison by looking at the scenario with $4$-bit quantization and no PCA, compared with PCA at $0.96$ variance retention and no quantization. They achieve a similar median F1-Score ($0.92$ and $0.90$ respectively), but PCA falls short of the quantization scenario: the former leads to a compression ratio of $155$ (thus increasing data size), while the latter allows compressing data to one-fifth size.

\begin{table}[t]
    \centering
    \setlength{\tabcolsep}{3pt}
    \vspace{3mm}
    \caption{Entropy for different PCA variance value}
    \label{tab:entropy_tab}
    \footnotesize
    \renewcommand{\arraystretch}{1.2}
    \begin{tabular}{r|c|c|c|c|c|c}
        $$$$ & Original &  \multicolumn{4}{c|}{PCA, at variance:}  \\
        $$$$ & $$$$ &  0.99 & 0.96 & 0.8 & 0.5  \\
        \hline
        Feature No. (median) &  175 & 115 & 100 & 80 & 15 \\
        Entropy (median) & 8.37 & 22.56 & 22.27 & 21.4 & 19.56 \\
        Non-Gzip Compressed [\%] & 611.40 & 219.28 & 182.78 & 109.12 & 40.74 \\
        Gzip Compressed [\%] & 100 & 188.27 & 155 & 90.42 & 29.74 \\
    \end{tabular}
    \vspace{-3mm}
\end{table}

Overall, these results indicate that scalar quantization without PCA achieves the best trade-off between compression ratio and data utility. PCA, although preserving data utility, spreads information across principal components, which increases entropy and reduces data compressibility. This leads to poor compression ratios, even increasing data size. Thus, for applications requiring both effective compression and high data utility, scalar quantization alone, without PCA preprocessing, appears to be a better approach.



\subsection{Vector Quantization}


\begin{figure*}[htbp] 
    \centering
    \begin{subfigure}[b]{0.30\linewidth}
        \centering
        \includegraphics[width=\linewidth]{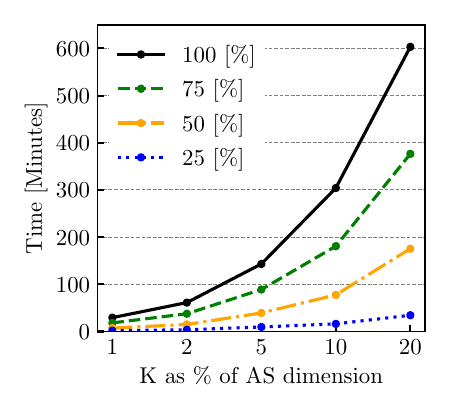}
        \caption{Computational Time}
        \label{fig:kmeans-times}
    \end{subfigure}
    \hfill 
    \begin{subfigure}[b]{0.30\linewidth}
        \centering
        \includegraphics[width=\linewidth]{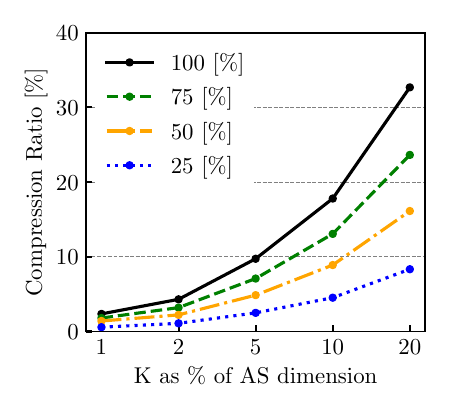}
        \caption{Compression Ratio}
        \label{fig:kmeans-compression}
    \end{subfigure}
    \hfill 
    \begin{subfigure}[b]{0.38\linewidth}
        \centering
        \includegraphics[width=\linewidth]{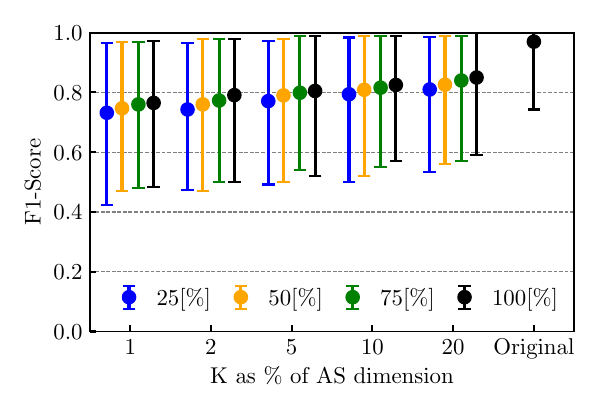}
        \caption{Classification Performance}
        \label{fig:kmeans-classification}       
    \end{subfigure}
    \vspace{-1mm}
    \caption{Comparative Analysis of Vector Quantization Through K-means}
    \vspace{-5mm}
    \label{fig:kmeans}
\end{figure*}

We now discuss results when using vector quantization, i.e., mapping entries to a discrete space by clustering similar data points. As a clustering algorithm, we adopt K-means as it proved to be the most scalable approach, while, our initial experiments using DBSCAN and HDBSCAN fell short because of impractical computational time. We evaluate the K-means clustering methodology on $30$ of the $35$ ASes datasets, with entries from \num{23100} to \num{715000}. Due to memory constraints -- we run experiments on a server equipped with \SI{192}{\giga\byte} RAM --, we are limited in testing ASes with significantly larger entries. Moreover, the computational time represents an additional issue as we show in the following.

In Figure~\ref{fig:kmeans}, we summarize our results in terms of computational time, compression ratio and F1-Score. In Figure~\ref{fig:kmeans-times}, we show the total time needed to compress data, running K-means with different $K$ values and then assigning each point to the corresponding centroid. The value of $k$ is set as a percentage of the entries within each AS (limited from $1$\% to $20$\%). The solid black line clearly shows that time is directly proportional to $K$, leading to a compression time of more than $603$ minutes when $K$ is set to achieve the finest data granularity considered -- $20$\% of the entries within each AS. Indeed, computational time increases linearly with \(K\) as defined in Equation~\ref{eq:kmeans_complexity}. Notice that running scalar quantization on the same data requires $19$ minutes only. To reduce computational time we also evaluate the performance of using a sub-sample ($25\%$, $50\%$ and $75\%$) of the original data (see the dashed lines). Reducing the dataset size to $50$\% lowers the computational time from $603$ to $175$ minutes with $K$ dimension of $20$\% the ASes. Additionally, increasing $K$ results in a proportional increase in time; for instance, looking at the solid black line the time rises from $143$ to $304$ minutes when increases $K$ dimension from $5$\% to $10$\%.

In Figure~\ref{fig:kmeans-compression}, we show the compression ratio results. As expected, the compression ratio depends directly on \(K\), since the compressed data consists of the list of centroids. With low \(K\), it is possible to achieve arbitrarily low compression ratios -- e.g., $2.36\%$ when $K$ is $1$\% the ASes. In Figure~\ref{fig:kmeans-classification}, we present the classification performance. Various combinations of \(K\) and AS dimension sub-samples reach the target of $0.80$ median F1-Score, representing a moderate yet sizable penalty -- the original F1-Score on the right-most label (``Original'') serves as a benchmark. Although increasing \(K\) generally improves the classification score, the gain is not substantial. Further larger \(K\) values may provide a better data representation and, likely, an F1-Score over to the target, albeit at the cost of increased computational time and worst compression ratio. Interestingly, running K-means on a subsample of the data has limited impact on F1-Score performance -- see the different error bars, referring to runs with different sampling ratios. 

In summary, these experiments show that vector quantization allows obtaining an arbitrary compression ratio by easily tuning the $K$ centroids number. However, this is achieved at the cost of high computational time required to cluster entries, and it entails a degradation of data utility. Sub-sampling the data used to obtain the clusters is a simple yet effective way to reduce computational time with minimal impact on data utility. 

\section{Conclusion and Future Work}
\label{sec:conclu}

This work evaluated the effectiveness of various lossy compression techniques for network flow records. Our findings highlight that the simplest method, Scalar Quantization, offers the best balance between compression efficiency and data utility. It achieves a tenfold reduction in storage size compared to entropic compression alone while maintaining the data utility for precise classification, with only a minimal impact compared to the original data. We analyzed more sophisticated methods as well, but they exhibited notable limitations. Specifically, Principal Component Analysis, although a popular dimensionality reduction tool, introduces significant entropy, rendering the transformed data unsuitable for further entropic compression. On the other hand, Vector Quantization, evaluated using the K-means clustering algorithm, suffers from severe scalability challenges and only becomes effective with an impractically large centroids number. Ultimately, our results affirm that simplicity, as emphasized by the ``Keep It Simple, Stupid'' (KISS) principle, is the most effective strategy in this context. Nevertheless, there remain broad avenues for future work. We plan to investigate advanced neural network-based approaches, particularly autoencoder architectures. They have shown potential in recent studies such as ~\cite{pekar2024autoflow}, even if on smaller datasets in terms of the number of flow records and features.


\section*{Acknowledgments}
The research leading to these results has been funded by projects 20228FT78M ``DREAM'' and 2022M2Z728 ``COMPACT'' under the Italian Ministry of University and Research 2022 PRIN program and by the Next Generation EU, Mission 4 Component 1 (CUP: D53D23001340006).

\nocite{*}
\bibliography{references}
\bibliographystyle{IEEEtran}

\end{document}